\documentclass[reprint,amsmath,twocolumn,amssymb,prb,superscriptaddress]{revtex4}
\usepackage{graphicx}
\usepackage{dcolumn}
\usepackage{bm}
\usepackage[utf8]{inputenc}
\usepackage{color}
\usepackage{hyperref}
\usepackage{siunitx}
\usepackage{amsmath}
\usepackage{ulem}
\usepackage{float}
\definecolor{purple}{RGB}{128,0,128}

\begin{document}

	\title{Heterostrain rules the flat-bands in magic-angle twisted graphene layers}

	\author{Florie Mesple}
	\email{email: florie.mesple@cea.fr; vincent.renard@cea.fr}
	\affiliation{Univ. Grenoble Alpes, CEA, IRIG, PHELIQS, F-38000 Grenoble, France}
	
	\author{Ahmed Missaoui}
	\affiliation{Laboratoire de Physique Théorique et Modélisation (UMR 8089), CY Cergy Paris Université, CNRS, 95302 Cergy-Pontoise, France}
	\author{Tommaso Cea}
	\affiliation{Imdea Nanoscience, Faraday 9, 28015 Madrid, Spain}
	\affiliation{Instituto de Ciencia de Materiales de Madrid, CSIC, Sor Juana Inés de la Cruz 3, Cantoblanco, 28049 Madrid, Spain}
	\author{Loic Huder}
	\affiliation{European Synchrotron Radiation Facility (ESRF), 71 Avenue des Martyrs, 38000 Grenoble, France.}
	\author{Francesco Guinea}
	\affiliation{Imdea Nanoscience, Faraday 9, 28015 Madrid, Spain}
	\affiliation{Donostia International Physics Center, Paseo Manuel de Lardizábal 4, 20018 San Sebastián, Spain}
	\author{Guy Trambly de Laissardière}
	\affiliation{Laboratoire de Physique Théorique et Modélisation, Université de Cergy-Pontoise-CNRS, F-95302 Cergy-Pontoise, France}
	
	\author{Claude Chapelier}
	\affiliation{Univ. Grenoble Alpes, CEA, IRIG, PHELIQS, F-38000 Grenoble, France}
	\author{Vincent T. Renard}
	\email{email: florie.mesple@cea.fr; vincent.renard@cea.fr}
	\affiliation{Univ. Grenoble Alpes, CEA, IRIG, PHELIQS, F-38000 Grenoble, France}


	\date{\today}
	
	
	
	\begin{abstract}
	 
	\end{abstract}
	
\maketitle

{\bf The moiré of twisted graphene bilayers can generate flat bands in which charge carriers do not posses enough kinetic energy to escape Coulomb interactions with each other\cite{Trambly2010,Bistritzer2011} leading to the formation of novel strongly correlated electronic states\cite{Cao2019,Cao2019b}. This exceptionally rich physics relies on the precise arrangement between the layers.\cite{Carr2018,Yankovitz2019}
We survey published Scanning Tunnelling Microscope (STM) measurements\cite{Huder2018,Kerelski2019,Choi2019,Jiangandrei2019,Xieyazdani2019,Zhang2020,Wong2019,Choi2020,Nuckolls2020,LinHe2020} to prove that near the magic angle, native heterostrain, the relative deformations between the layers, dominates twist in determining the flat bands. This is demonstrated at large doping where electronic correlations have a weak effect and where we also show that tip-induced strain can have a strong influence. In the opposite situation of low doping, we find that electronic correlation further normalize the flat bands in a way that strongly depends on experimental details.}

The strongly correlated electron physics recently observed in twisted graphene layers\cite{Cao2019,Cao2019b} develops in flat bands \cite{Trambly2010,Bistritzer2011} which are very sensitive to the relative arrangement between the layers. For instance, the superconducting phase has been reported to occur in a very narrow range of rotation angle between the layers around the magic angle. As another illustration, hydrostatic pressure changes the interlayer distance which also strongly influences superconductivity \cite{Carr2018,Yankovitz2019}.  Heterostrain, the in-plane deformation of one layer with respect to the other is an ubiquitous source of modification of the relative arrangement between the layers\cite{Huder2018,Kerelski2019,Choi2019,Jiangandrei2019,Xieyazdani2019,Zhang2020,Wong2019,Choi2020,Nuckolls2020,LinHe2020}. Experiments\cite{Huder2018} and theory\cite{Falko2014,Bi2019designing} have shown that heterostrain affects the flat bands which could have an impact on the strongly correlated electron physics. Bi, Yuan and Fu have even predicted that near the magic angle, the bandwith becomes {\it insensitive} to the twist angle and that the effect of heterostrain is {\it dominant},\cite{Bi2019designing} calling for systematic experimental study of its effect. Such program is however difficult to implement owing to the lack of controllability of heterostrain, and its inhomogeneity previously referred to as "twist angle disorder" and which was also found to impact strongly correlated states at the macroscopic scale \cite{Lu2019,Choi2020}. In order to overcome these issues, we survey already published experimental STM data\cite{Huder2018,Kerelski2019,Choi2019,Jiangandrei2019,Xieyazdani2019,Wong2019,Zhang2020,Choi2020,Nuckolls2020} in view of quantifying the effect of homogeneous heterostrain on the physics of magic-angle twisted graphene layers.
\begin{figure*}[]
	\includegraphics[width=\textwidth]{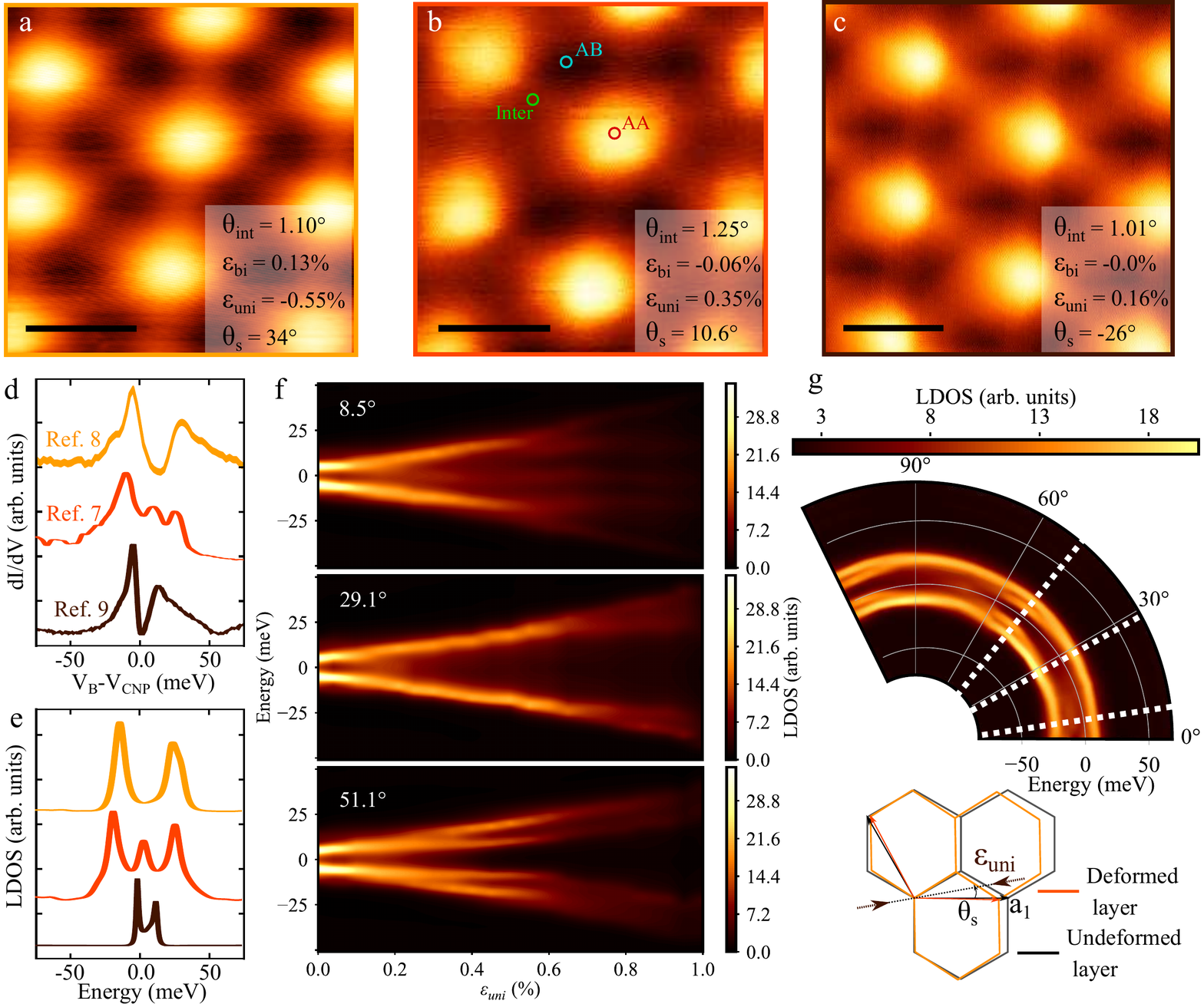}
	\caption{{Heterostrained twisted graphene layers.} Panels \textbf{a}, \textbf{b} and \textbf{c} present STM images of twisted graphene layers near the magic angle adapted from Refs~\cite{Kerelski2019}, \cite{Huder2018} and \cite{Choi2019} respectively. The scale bar is 10 nm in each image. Insets present the parameters describing the relative arrangement of the layers in good agreement with estimates of the original studies.\cite{Huder2018,Kerelski2019,Choi2019} \textbf{d} Local density of states measured in the AA regions for each samples of panels \textbf{a}, \textbf{b} and \textbf{c}. \textbf{e} Corresponding tight-binding calculation of the LDOS including heterostrain.  \textbf{f} Tight-binding prediction for the local density of states in AA regions as function of energy and for increasing uniaxial heterostrain. The variations are plotted for three different angles $\theta_s$ of application of heterostrain. \textbf{g} Local density of states calculated for 0.4~\% of heterostrain applying along varying $\theta_s$. The white dotted lines indicate the values of angles used in panel \textbf{f}. \textbf{h} Sketch of uniaxial heterostrain configuration. One layer is deformed by a uniaxial heterostrain applied along the direction defined by $\theta_s$ and then rotated by and angle $\theta$ with respect to the undeformed layer.}
	\label{fig:Data}
\end{figure*}

Figure~\ref{fig:Data} presents typical STM images collected from Refs.~\onlinecite{Huder2018,Kerelski2019,Choi2019}. An immediate observation is that the STM images all look very similar. This is not surprising because these samples have a twist angle very close to one another. This similarity is only apparent as evidenced by the variety of shapes of the local density of states (LDOS) measured from the dI/dV(V) spectroscopy (Fig.~\ref{fig:Data}d). Despite the flat-bands should merge at the magic angle, the spectra of Ref.~\onlinecite{Kerelski2019} and \onlinecite{Choi2019} show two van Hove singularities indicating that the flat bands are still separated. Their spacing $\Delta E_{exp}$ is doping dependent which has been attributed to electron-electron interactions.\cite{Kerelski2019,Choi2019,Jiangandrei2019,Xieyazdani2019,Wong2019,Zhang2020} It can reach several tens of meV, even at large doping where correlations are not expected to renormalize strongly the bands. Still, while the two samples have a twist angle differing by only 0.1$^\circ$, doping does not explain why their $\Delta E_{exp}$ differ by a factor 2-3.  Sample to sample variation is obvious from other published data.\cite{Jiangandrei2019,Xieyazdani2019,Wong2019,Zhang2020,Choi2020,Nuckolls2020} Strikingly the data of Ref.~\onlinecite{Huder2018} show a third peak at zero energy, a result which was recently reproduced\cite{LinHe2020}. In order to determine whether this variety in sample behaviour can be understood within the framework of heterostrain, we use the method described in Ref.~\onlinecite{Artaud2016} to determine the precise arrangement of the layers and calculate the corresponding local density of states using a tight-binding method.\cite{Trambly2010,Brihuega2012,Trambly2012b}

\begin{figure*}[]
	\centering
	\includegraphics[width=0.8\textwidth]{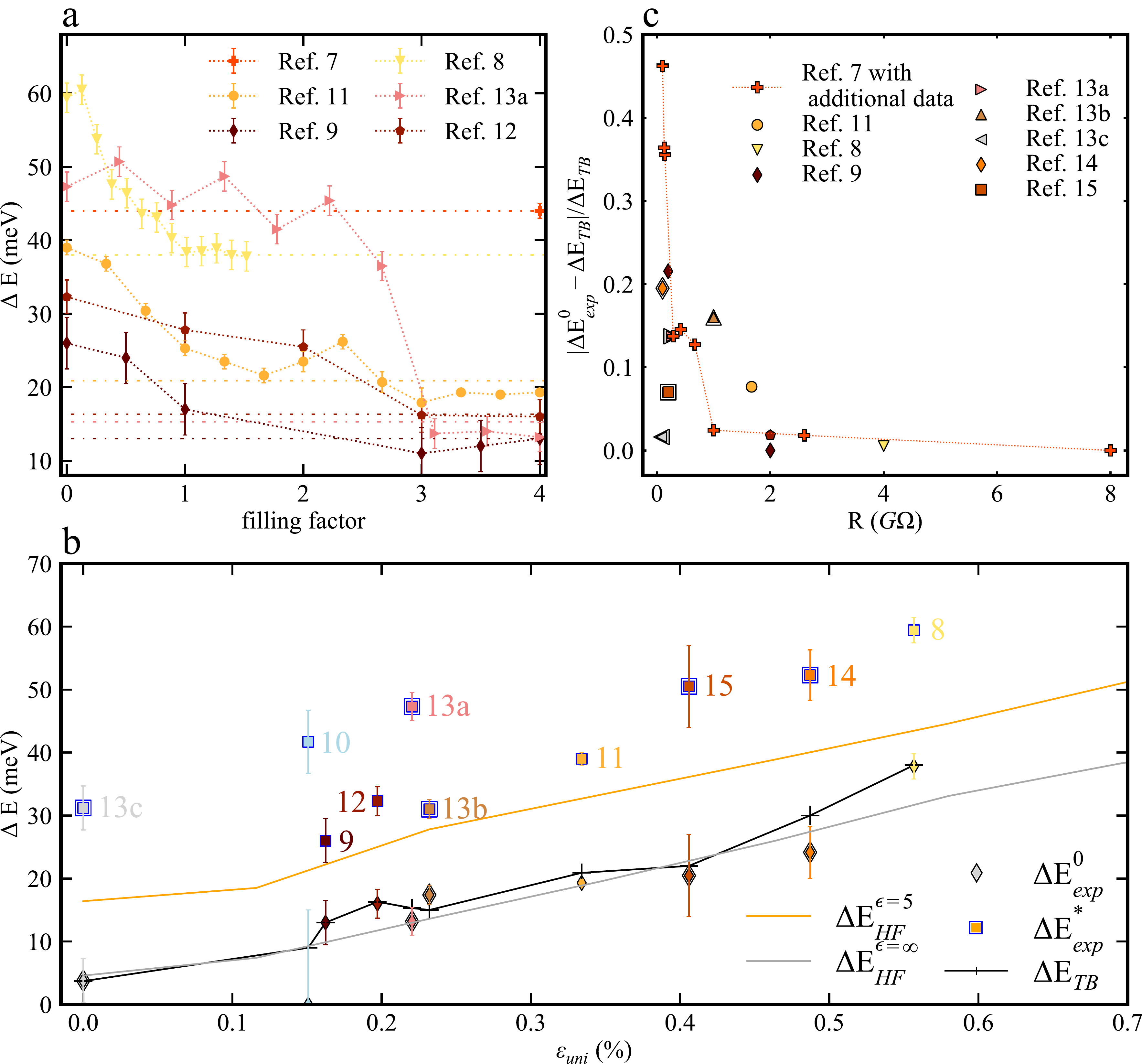}
	\caption{{Comparison between theory and experiments}. \textbf{a} Representative doping dependence of the spacing between van Hove singularities. The horizontal line corresponds to prediction by tight-binding calculations with heterostrain. \textbf{b} Experimental spacing of van Hove singularities at large doping ($\Delta E_{exp}^{0}$), zero doping ($\Delta E_{exp}^{*}$), theoretical tight-binding ($\Delta E_{TB}$), low-interaction Hartree-Fock ($\Delta E_{HF}^{\epsilon=\infty}$) and large interaction $\Delta E_{HF}^{\epsilon=5}$.  Such presentation of all the data on a single graph is justified by i) the weak dependence of $\Delta E_{exp}$ on the twist angle near the magic angle (See Ref.~\cite{Bi2019designing} and supplementary Fig.~S5) and ii) the weak dependence of $\Delta E$ on $\theta_s$ if it is estimated as the spacing between the outermost singularities (Fig.~\ref{fig:Data}g). This is also why the Hartree-Fock model was calculated for a twist angle of 1.1$^\circ$ and $\theta_s=30^\circ$. On the contrary, $\Delta E_{TB}$ were calculated with the full relative arrangement of each experimental data and show a better agreement with experiments. The twist angle has still a small influence which explains the small deviations from a purely linear strain dependence. The measurements showing a cascade of transitions are indicated by a double border. \textbf{c} Relative deviation of $\Delta E_{exp}^{0}$ to $\Delta E_{TB}$ as a function of the tunnelling resistance  $R_t=V_b/i_t$. The figure includes new data for the sample of Fig.~\ref*{fig:Data}b.}
	\label{fig:Theory vs experiments}
\end{figure*}

Figure.~\ref{fig:Data}a, b and c include our estimates of heterostrain. In all cases, biaxial heterostrain is smaller than uniaxial heterostrain which varies by a factor 3 from the smallest to the largest. Figure.~\ref{fig:Data}d shows that the calculated local density of states agrees very well with experimental data. In particular the number of peaks is controlled by heterostrain which also influences their spacing pointing to a strong contribution of heterostrain to the observed phenomenology. The tight-binding calculations presented in Fig.~\ref{fig:Data}f show that the van Hove singularities separate linearly with uniaxial heterostrain for all angle $\theta_s$ of application of heterostrain. This angle controls the splitting of the van Hove singularities (Fig.~\ref*{fig:Data}g) leading to three typical behaviours reported in Fig.~\ref{fig:Data}f  (See supplementary information and videos for more theoretical results). These results agree with those of the continuum model \cite{Bi2019designing}. The more quantitative comparison of Fig.~\ref{fig:Theory vs experiments}a shows that $\Delta E_{exp}$ converges to the value $\Delta E_{TB}$ predicted by our tight-binding calculation at large doping. This has to be expected because the effect of electronic correlations measured by the ratio of Coulomb to kinetic energy reduces with doping and the system evolves towards the non-interacting situation modeled by our tight binding calculations. It establishes that heterostrain controls the physics of twisted graphene layers near the magic angle at large doping. This can be viewed explicitly in Fig.~\ref{fig:Theory vs experiments}b which shows that $\Delta E_{exp}^0$, the experimental spacings of van Hove singularities at large doping, depends linearly on heterostrain as predicted by tight binding and the continuum model. On the contrary, the same data plotted as function of the twist angle do not show particular correlation with this parameter (Supplementary Fig.~S5).

\begin{figure*}[t]
	\includegraphics[width=\textwidth]{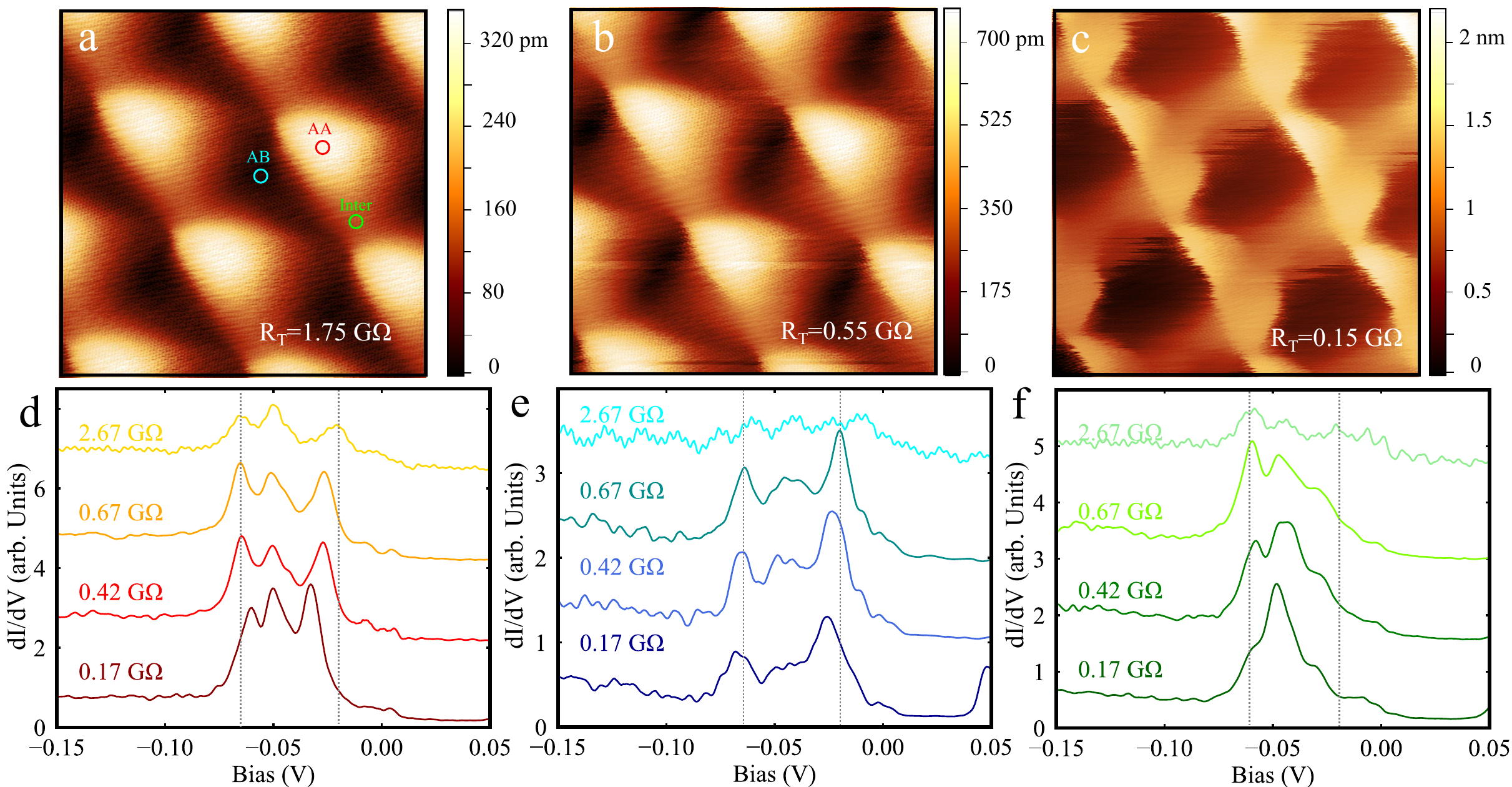}
	\caption{{Tip-induced strain}. \textbf{a, b, c} STM images measured at decreasing tunnel resistance. The tunnelling set point is $i_t=100$ pA and the bias is $V_b=175$ mV in \textbf{a}, $V_b=55$ mV in \textbf{b} and $V_b=15$ mV in \textbf{c}. The images are 25$\times$25 nm$^2$. \textbf{d, e, f} Local density of states measured by STM for decreasing tunnelling conductance in AA (\textbf{d}), AB (\textbf{e}) and intermediate regions (\textbf{f}). The vertical dotted lines are guides for the eyes to track the position of the LDOS peaks.}
	\label{fig:Tip induced strain}
\end{figure*}

Figure~\ref{fig:Theory vs experiments}c provides a deeper level of comparison between theory and experiment. It presents the relative difference between the $\Delta E_{TB}$ and the experimental $\Delta E_{exp}^{0}$ as function of the tunnelling resistance $R_t=V_b/i_t$ ($V_b$ and $i_t$ are the tunnelling bias and current). The excellent agreement obtained for $R_t>$  2 G$\Omega$ degrades below this value pointing to a possible influence of tip-induced strain which is known to be controlled by $R_t$.\cite{Mamin1986} The deformations seen in Fig~\ref{fig:Tip induced strain}a, b and c showing the evolution of the image of Fig.~\ref{fig:Data}b for decreasing $R_t$ corroborate this interpretation (The reader may also refer to Fig.~S3 of Ref.~\cite{Choi2019} for another example). 

Figure~\ref{fig:Tip induced strain} shows that while spatial variations of the LDOS at high $R_t$ only reflect electronic localisation in AA regions, the response to tip-induced strain at low $R_t$ strongly depends on the position on the moiré pattern. While AB regions are weakly affected (Fig.~\ref{fig:Tip induced strain}e), $\Delta E_{exp}$ is reduced by 40\% at low tunnel resistance in AA regions indicating a flattening of the bands (Fig.~\ref{fig:Tip induced strain}d). The flattening culminates in intermediate regions between two AA regions where a 15 meV wide single LDOS peak signals that very flat bands can be engineered there (Fig.~\ref{fig:Tip induced strain}f). It is not surprising that these regions are the most sensitive to tip-induced strain because they are characterized by an excess of elastic energy \cite{Alden2013,Wijk2015,Yoo2019,Gornostyrev2020} and are therefore more easily perturbed. We note that the tip may induce a complex strain pattern including hetero-, homostrain as well as vertical displacement all of which depend on the position on the moiré. It follows that the relative arrangement between the layers can no longer be determined from STM images at low $R_t$ and that the response to tip-induced strain is largely sample dependent. This is illustrated by Fig.~S5 of Ref.~\onlinecite{Choi2019} in which, contrary to our experiment, the spacing between van Hove singularities increases with decreasing $R_t$. We conclude that great care is needed when discussing quantitatively experimental results and that the STM tip can be used to locally engineer the relative arrangement between the layers.

The very good agreement between experiments and theory obtained so far raises the question about the influence of heterostrain on electronic correlations which express most at zero doping.\cite{Kerelski2019,Choi2019,Jiangandrei2019,Xieyazdani2019,Wong2019,Zhang2020} Figure~\ref{fig:Theory vs experiments}b shows that $\Delta E_{exp}^*$, the experimental spacing of van Hove singularities in this regime is increased by electronic correlations and tends to increase with strain. This tendency is well pictured by our Hartree-Fock calculations presented in Fig.~\ref{fig:Theory vs experiments}b including both heterostrain and electronic correlations for the dielectric constant $\epsilon= 5$ (See Ref.~\onlinecite{Cea2019} and supplementary information for the details). These calculations however underestimate $\Delta E_{exp}^*$. Further increasing the effect of interactions by decreasing the dielectric constant does not lead to a better agreement especially at large strains where the low energy bands become so wide that they merge into the continuum of the higher energy bands and van Hove singularities can no longer be clearly defined (Supplementary information). Also the calculations do not capture the large experimental scatter of  $\Delta E_{exp}^*$ which we were not able to correlate to any experimental parameter (twist angle, heterostrain value or angle of application, bandwidths as measured by the FWHM of van Hove singularities, tunnelling resistance, temperature). The scatter also does not correlate with the appearance of polarized states leading to the cascade of transitions seen in some samples at intermediate fillings.\cite{Wong2019,Choi2020,Nuckolls2020} This points to a strong sensitivity of electronic correlations to some additional experimental parameters beyond those investigated here. This could be due to the substrate or more generally to the detailed electrostatic environment as suggested by several studies reviewed in Ref. \onlinecite{balents_superconductivity_2020}, and to atomic lattice relaxation effects \cite{Yoo2019} calling for a systematic experimental study of the effect of those parameters.

Returning to heterostrain, its strong impact on the flat bands of magic angle twisted graphene layers also calls for a systematic investigation of its influence on the strongly correlated phases and that of other moiré materials. In this context, it would be extremely desirable to be able to tune it. Alternatively, one could also rely of the variability in the fabrication process to generate a representative set of samples such as the one we have studied here.

\bibliographystyle{nature.bst}
\bibliography{Biblio_Heterostrain.bib}

\section*{Methods}
\subsection*{Data collection}
When possible, the raw data used in the survey were retrieved from the authors of the publications. Otherwise, the data were digitized from the corresponding publications. 
We could use the method of Ref.~\cite{Artaud2016} to extract heterostrain for Refs.~\cite{Huder2018,Kerelski2019,Choi2019} which had STM images with atomic resolution. The strength of this method is to isolate heterostrain from homostrain and possible artefacts due to the calibration of the piezoelectric tube. For Refs. \textit{(10, 11, 13a, 14, 15)}, we used the method described in Ref.~\cite{Kerelski2019} to determine heterostrain. Although it does not allow to determine biaxial heterostrain we found it to provide good estimates of uniaxial heterostrain for the data of Refs.~\cite{Huder2018,Kerelski2019,Choi2019}. For Refs. \textit{(12, 13b, 13c)}, we did not have the data to do our own analysis and used the values of heterostrain provided by the authors. Error bars in Fig. 2 of main text present the uncertainty as estimated the full width at half maximum of a van Hove singularity.

\subsection*{STM measurements}
We performed the measurements presented in Figure 3 of the main text in order to document the effect of tip-induced strain on the twisted graphene layers described in Ref.~\cite{Huder2018}. Tip-induced strain was changed by changing the bias for a given tunnel current in imaging conditions. In spectroscopic mode, the tip-sample interaction was defined by the set point prior to switching off the feedback loop and subsequent sweeping of the bias voltage between -200 mV and 200 mV. The $dI/dV(V)$ signal was measured using phase sensitive detection with a 2 mV oscillation at 263 Hz added to the tunnel bias. The di/dV curves were normalised to 1 at -200 mV.


\section*{Acknowledgements}

The authors thank E. Andrei, S. Nadj-Perge, for sharing their data and for interesting discussions. AM is supported by Paris//Seine excellence initiative (Grant No. 2019-055-C01-A0). FG and TC acknowledge funding from the European Commision, under the Graphene Flagship, Core 3, grant number 881603, and from grants NMAT2D (Comunidad de Madrid, Spain),  SprQuMat and SEV-2016-0686, (Ministerio de Ciencia e Innovación, Spain) 

\section*{Author contributions:}
VTR,FM and CC defined the project. FM performed the data analysis from the literature and provided the cells for tight binding calculations which were performed by AM and GTL. TC and FG performed Hartree-Fock calculations. LH, VTR and CC performed the STM experiments. VTR and FM wrote the manuscript with input from all the authors. VTR supervised the project.

\section*{Supplementary materials}
Supplementary file and videos available on request to vincent.renard@cea.fr.

\clearpage

\end{document}